\documentclass[12pt]{iopart}
\usepackage[utf8]{inputenc}
\usepackage{iopams}
\usepackage{graphicx}
\usepackage{amssymb}
\usepackage{epstopdf}
\usepackage{subfigure}
\usepackage{float}

\usepackage{iopams}

\begin{document}
\title[Network of Surnames]{The complex social network of surnames: A comparison between Brazil and Portugal}
\author{G. D. Ferreira$^{1}$, G. M. Viswanathan$^{1,2}$, L. R. da Silva$^{1,2}$ and H. J. Herrmann$^{2,3}$}
\address{$^1$ Universidade Federal do Rio Grande do Norte, Departamento de F\'isica Te\'orica e Experimental, Natal-RN, 59078-900, Brazil.}
\address{$^2$ National Institute of Science and Technology of Complex Systems, Brazil.}
\address{$^3$ Institut f\"{u}r Baustoffe (IfB), HIF E 12, Stefano-Franscini-Platz 3,
8093 Z\"{u}rich.}
\ead{gerda.duarte@dfte.ufrn.br, gandhi@dfte.ufrn.br, luciano@dfte.ufrn.br, hans@ifb.baug.ethz.ch}
\vspace{10pt}
\begin{indented}
\item[]May 2017
\end{indented}

\begin{abstract}
We present a study of social networks based on the analysis of Brazilian and Portuguese family names (surnames). We construct networks whose nodes are names of families and whose edges represent parental relations between two families. From these networks we extract the connectivity distribution, clustering coefficient, shortest path and centrality. We find that the connectivity distribution follows an approximate power law. We associate the number of hubs, centrality and entropy to the degree of miscegenation in the societies in both countries. Our results show that Portuguese society has a higher miscegenation degree than Brazilian society. All networks analyzed lead to approximate inverse square power laws in the degree distribution. We conclude that the thermodynamic limit is reached for small networks (3 or 4 thousand nodes). The assortative mixing of all networks is negative, showing that the more connected vertices are connected to vertices with lower connectivity. Finally, the network of surnames presents some small world characteristics.
\end{abstract}
\pacs{64.60.aq, 89.75.-k, 89.75.Da}


\vspace{2pc}
\noindent{\it Keywords}: Complex Network, Family Names, Centrality, Assortative Mixing, Entropy
 
\submitto{\JSTAT}

\maketitle


\section{Introduction}
\qquad The study of social networks typically involves analyzing the relations between people or organizations, as well as combinations of individuals groups and their interrelations \cite{GARTON1997} and \cite{HANNEMAN2001}. Mapping these relationships results in a complex network. The quantitative study of social networks uses theoretical and experimental tools from social sciences \cite{DURKHEIN1995}, \cite{MORENO1934}, mathematics (graph theory) \cite{HARARY1969} and physics \cite{DOROGOVTSEV2002}. Networks have two basic components: nodes (sites or vertices) and connections (edges or links) between them. For example, consider the spread of information (e.g., gossip) in a social network \cite{LIND2007} and \cite{LIND2007A}. A node then represents a person and the social relations between two people are mapped by the edges. A common assumption is that the more connected the people, the faster the information percolates over the network.\\ 
\qquad The study of these complex social networks is motivated by the understanding that many real systems do not follow classical distributions (e.g., Poisson or Gaussian). In this work we study the surnames of people, obtained from lists of persons who are related to each other by common specific characteristics, such as, for example, a phone list, a list of employees of a company, a freshmen list from a university, and so on. Our study starts with the premise that a large list in some way represents the society in which we live (or at least part of it). We analyze the frequency of these surnames and build the networks and the corresponding histograms of connectivities. Some work already has been done in this direction. For instance, in ref. \cite{MIYAZIMA2000}, the authors conducted a frequency study of Japanese families names and as a result, they found that the names are distributed according to a power law. They found that only a few family names appear very frequently, whereas most names occur rarely.

\section{Theoretical models}
\qquad Network definition depends on what questions one is interested in answering. An edge may represent the friendship between individuals, or represent professional relations, the exchange of goods or money, communication patterns, romantic relationships, or many other types of connection. One of the basic characteristics of a complex network is the distribution of the degree of its nodes, because this defines the structure of the network. The connectivity of a node $k$ is the number of edges connected to it. Then $p_k$ is the fraction of nodes in a network that has connectivity $k$. The value $p_k$ can also be thought of as the probability that a randomly chosen node on the network has connectivity $k$ ~\cite{NEWMAN2010}. If the degree distribution is a power law, we have:
\begin{equation}
p_k = Ck^{-\gamma},
\label{eq-pk}
\end{equation}
where $C$ is a constant used to normalize the distribution. Many networks have a distribution of connectivity that approximately follows a power law \cite{BARABASI2002}, \cite{NEWMAN2003a} and \cite{NEWMAN2003b}.
In many cases it is useful to consider also the complementary cumulative distribution function or CDF of a power-law distributed variable, which we denote by $P_k$ \cite{CLAUSET2009}.
\begin{equation}
P_k = \int_{k}^{\infty} p(k')dk'.
\label{eq-cdf-Pk} 
\end{equation}
\qquad Price \cite{PRICE1965} proposed a simple and elegant model of network formation to study the network of citations that gives rise to a connectivity distribution power law. The model proposed by Price is a growing network that has an exact solution, showing that the distribution of connectivity has a tail with a power law behavior. Barab\'asi and Albert \cite{BARABASI1999} and \cite{BARABASI_ALBERT2002} proposed a growing network model (along with \textquotedblleft{}preferential binding\textquotedblright{} expression). This model, which is certainly the best known complex network model nowadays, is similar to Price, but not identical. In the model of Barab\'asi-Albert, the higher the connectivity of a node, the more likely would this node receive a new connection, which directly implies that the first nodes to be born on the network will become the hubs and that these can not be overcome by younger nodes. In real networks, that may not happen, for example the first company located in a city or state will not necessarily become an industrial pole, without another company exceeding it. In 2001, Barab\'asi and Bianconi added an ingredient called \textquotedblleft{}quality\textquotedblright{} (or fitness), in order to make the model closer to several real networks. Here \cite{BARABASI2001a} a greater number of nodes appears with high connectivity, although most of the nodes still possess low connectivity, i.e., this model has more hubs than the previous one. The model is based on the idea of quality bringing competitiveness to the nodes that is able to affect the evolution of the network. According to this idea, the intrinsic ability of the nodes to attract links in the network varies from node to node. The most efficient node (big quality) is able to receive more connections than the others. Accordingly, not all the nodes are identical to each other, each one is born with an proper ability. The preferential binding probability is given by:

\begin{equation}
\Pi_i = \frac{\eta_ik_i}{\sum_{j}\eta_jk_j},
\label{eq-bianconi-quality}
\end{equation}
where $\eta$ is the fitness of site $i$ and is assigned randomly.\\
\qquad The main properties of complex networks are the clustering coefficient $C$ and the shortest average path $\ell$.
The usual definition for the local clustering coefficient is as follows: Given a node $i$ then
\begin{equation}
C_i = \frac{\textrm{number of connections between neighbors of i}}{\textrm{total number of possible connections}},
\label{eq-clustering-coefficient}
\end{equation}
and the average clustering coefficient is given by:
\begin{equation}
C_m = \frac{\sum{C_i}}{N},
\label{eq-average-clustering-coefficient}
\end{equation}
where $N$ is the size of the network. The \textit{geodesic path} is the shortest path between two nodes over the network. We define $d_{ij}$ as the length of a geodesic path from node $i$ to node $j$. The geodesic distance from $i$ to $j$, averaged, over all nodes $j$ in the network, is given by:
\begin{equation}
\ell_i = \frac{1}{n-1}\sum_{j(\neq i)}d_{ij}.
\label{eq-lowest-path} 
\end{equation} 
The shortest average path is given by:
\begin{equation}
\ell = \frac{\sum{\ell_i}}{N}.
\label{eq-lowest-average-path}
\end{equation}


\section{Method}
\begin{figure*}[!htb]
\begin{center}
\includegraphics[scale = 0.45]{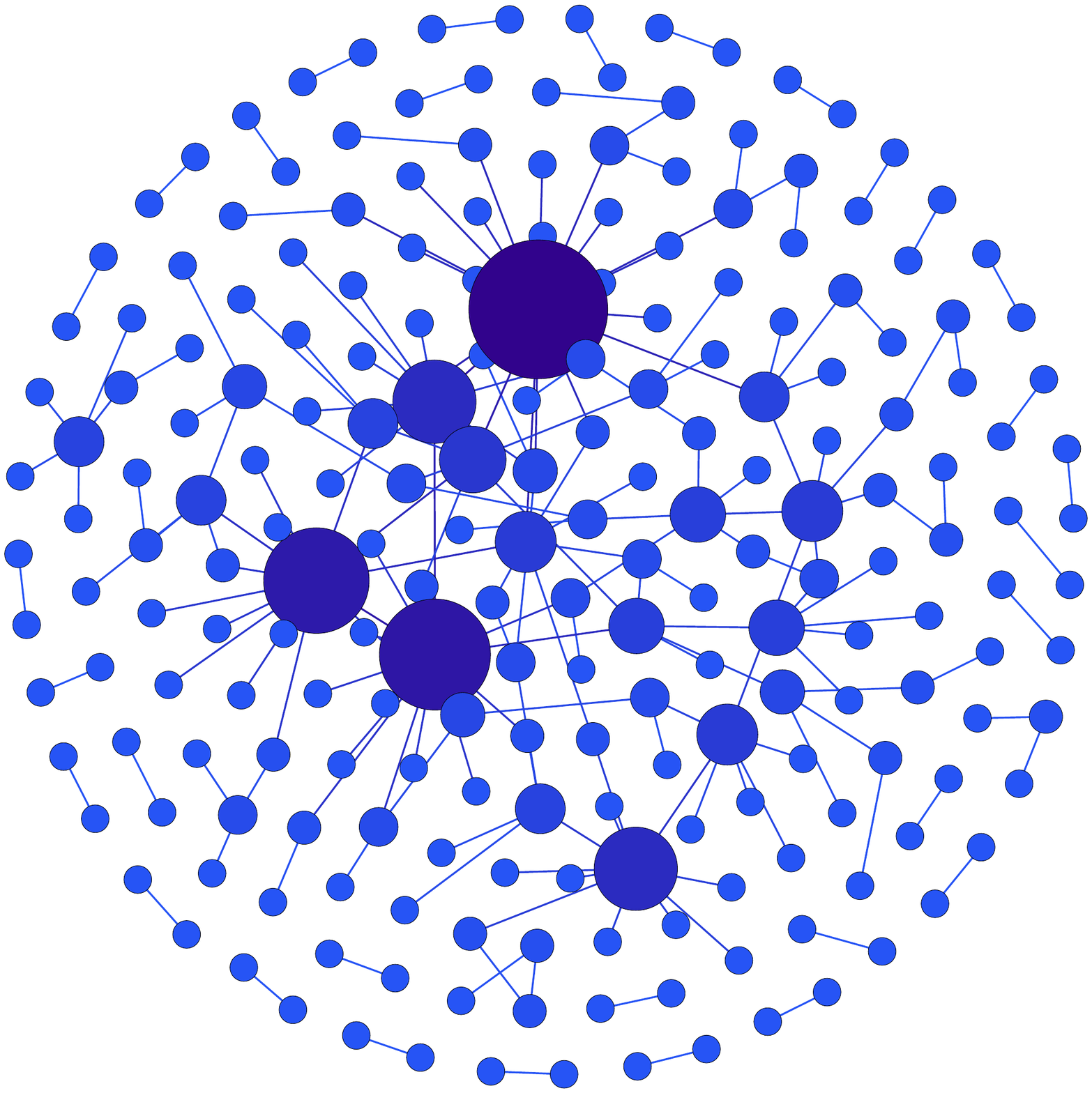}
\caption{\small Network of surnames built for a small sample of 250 people of Portuguese society, where we use the word \textbf{Jos\'e} to generate the sample. The vertex size is proportional to connectivity.} 
\label{Rede_amostra}
\end{center}
\end{figure*}

\begin{figure*}[!htb]
\begin{center}
\includegraphics[scale = 0.45]{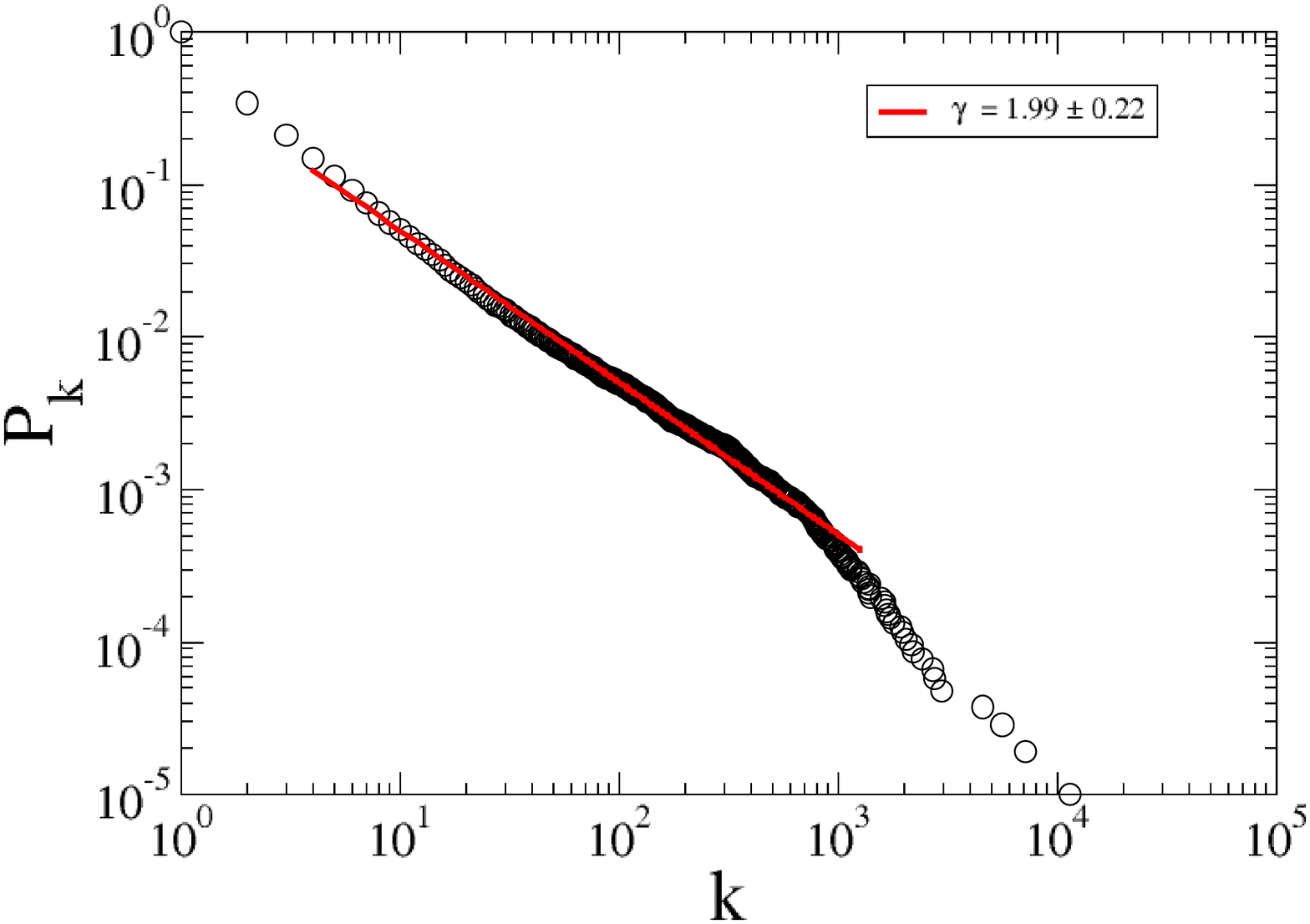}
\caption{\small The cumulative density functions (CDF) $P_k$ according to eq.~\ref{eq-cdf-Pk} for a sample with 104323 nodes of the Brazilian population whose names contain the word \textbf{Maria}. The plot shows the power law with $\gamma \simeq 2$ for the probability density $P_{k}$ vs. the connectivity $k$.} 
\label{Rede_maria}
\end{center}
\end{figure*}

\begin{figure*}[!htb]
\begin{center}
\includegraphics[scale = 0.45]{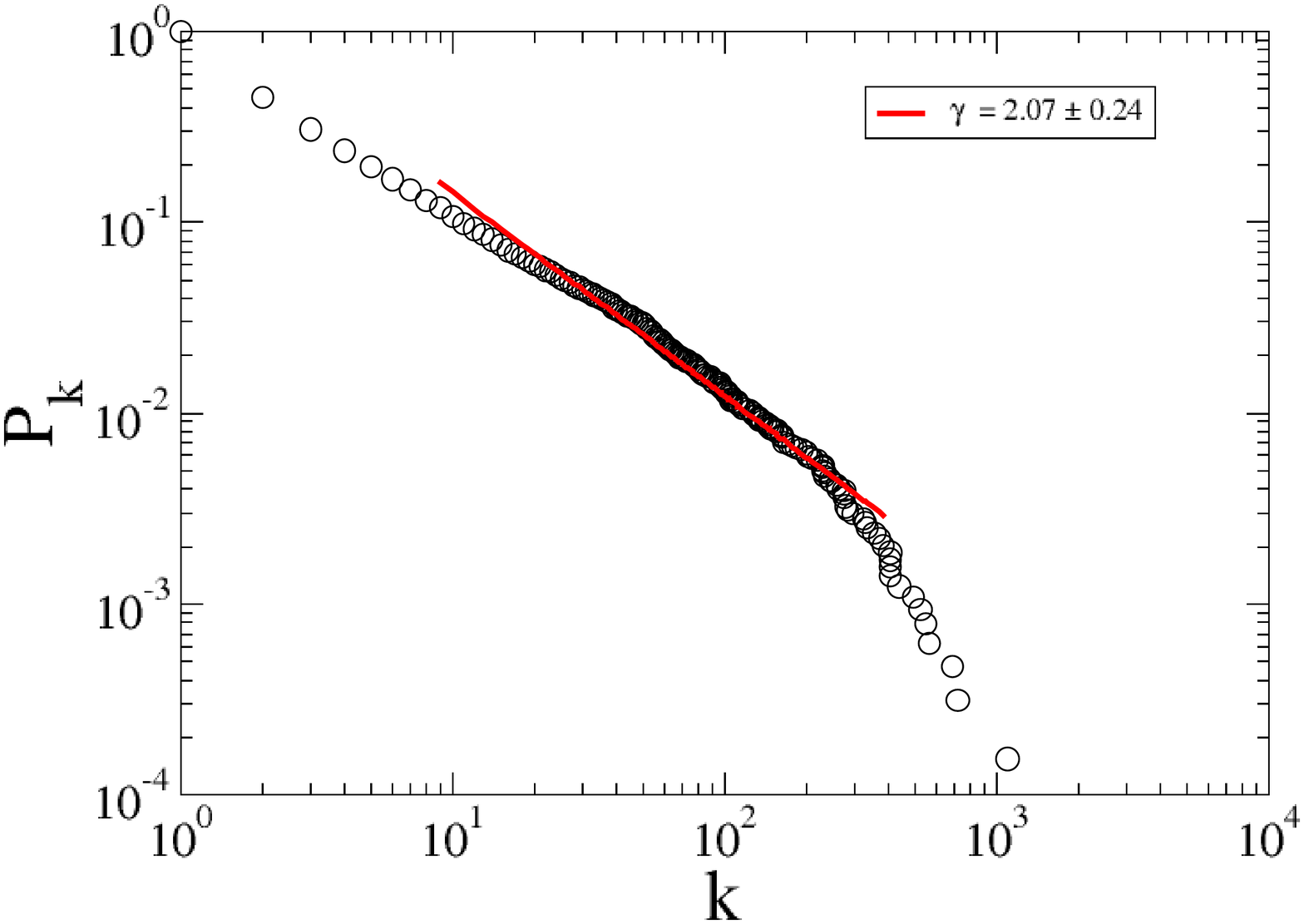}
\caption{\small The cumulative density functions (CDF) $P_k$ according to eq.~\ref{eq-cdf-Pk} for a sample with 6392 nodes of the Portuguese population whose names contain the word \textbf{Jos\'e}. The plot shows the power law with $\gamma \simeq 2$ for the probability density $P_{k}$ vs. the connectivity $k$.} 
\label{Rede_jose}
\end{center}
\end{figure*}

\qquad We build the networks from telephone listings using as a reference a first name (for example, Ana, Maria, Jo\~{a}o, etc.). We consider any surname as a node of the network. If we consider two nodes of the network, the corresponding edge is active if there is a person with surnames in both nodes. In the formation of the names in Brazil and Portugal it is common to use first the last family name of the mother and then the last family name of the father in the name of the child. In Fig.~\ref{Rede_amostra} we show a network for a sample with 250 people. In Fig. \ref{Rede_maria} we show the connectivity distribution for network built from the word \textbf{Maria} from Brazil and in Fig. \ref{Rede_jose} for the network built from the word \textbf{Jos\'e} from Portugal, i.e., all individuals who had the word Maria or Jos\'e in their names and that belonged to more than one family participated in this network. In the Barab\'asi-Albert model all nodes are connected while in the network of surnames there is the several disconnected clusters of nodes. Therefore, the network of surnames (or simply \textbf{NS}) is modeled by a disconnected graph.

\subsection{Algorithm}

$1-$ Our samples are obtained from telephone directories available on the Internet from a first name used for reference, for example, the network Ana\_BR is generated from the Brazilian phone list by collecting all the names that contain the word Ana;\\
$2-$ The listing is then treated to remove names that are considered invalid, such as names of hospitals, companies, drugstores, etc. and only people's full names remain;\\ 
$3-$ With the treated samples, we calculated statistical information like histograms of surnames (these surnames identify families, such as the Ferreira family, the Silva family, the Santos family, and so on);\\
$4-$ We only consider the last two family names of each person to build the network (if a person has three family names in their name, the last two surnames are used). To belong to the network of surnames it is necessary that an individual belongs to at least two families - The node enters in the network with at least one connection. For example, in the list formed by the word \textbf{Maria}, we have:
\begin{itemize}
\item Maria Carla de Souza Cavalcante;
\item Jo\~{a}o Maria Rocha;
\item Ana Maria Alencar de Souza;
\item ...
\end{itemize}
\qquad Maria Carla has two family names, so their families (Souza and Cavalcante) are vertices in the network because Maria Carla represents an edge connecting these two families. Jo\~{a}o Maria has only a family name (Rocha), so he does not participate in the network, because Jo\~{a}o Maria does not connect the Rocha family with no any other family. Ana Maria has two family names (Alencar and Souza), so she belongs to the network, because Ana Maria connects these two families. In the network, the Souza family receives a connection through Maria Carla with the Cavalcante family, and receives a new connection through Ana Maria with the Alencar family;\\ 
$5-$ We construct the network forming the adjacency lists, which in turn are a simplification of the adjacency matrix;\\
$6-$ With the network completed we calculate several properties like clustering coefficient, shortest average path, connectivity of each node, centrality, etc..


\section{Results}
\qquad In table \ref{tab: BRPT} we see the main information obtained from Brazilian and Portuguese lists.\\
\begin{table}
\caption{\label{tab: BRPT}\textit{Main characteristics of networks of surnames}. CT means the country; \textbf{$N$} represents the number of nodes in the network; \textbf{$k_h$} is the degree of the hub; \textbf{$C_m$} is the average clustering coefficient of the network; \textbf{$\ell$} is the shortest average path between any two nodes in the network; $\gamma$ is the exponent of the power law and was obtained by linear regression; \textbf{$Hub_c$} is the vertex with larger Closeness centrality; \textbf{$C_c$} is the value of Closeness centrality for this vertex; \textbf{$Hub_b$} is the vertex with larger Betweenness centrality; \textbf{$C_b$} is the value of Betweenness centrality for this vertex and $r$ is the Assortative mixing.}
\footnotesize
\begin{tabular}{@{}lllllllllllll}
\br
CT&network&$N$&Hub&$k_{h}$&$C_{m}$&$\ell$&$\gamma$&$Hub_{c}$&$C_{c}$&$Hub_{b}$&$C_{b}$&$r$\\
\mr
  BR & Ana & 22985 & Silva & 2334 & 0.07 & 3.73 & 1.88 $\pm$ 0.31 & Ribeiro & 0.44 & Silva & 0.14 & -0.20 \\  
  BR & Francisco & 13263 & Silva & 1570 & 0.06 & 3.67 & 1.92 $\pm$ 0.29 & Souza & 0.47 & Francisco & 0.16 & -0.23 \\
  BR & Jo\~{a}o & 17431 & Silva & 1803 & 0.09 & 3.63 & 1.96 $\pm$ 0.28 & Souza & 0.46 & Silva & 0.13 & -0.25 \\                                        
  BR & Jos\'e & 38651 & Silva & 4462 & 0.08 & 3.65 & 1.93 $\pm$ 0.26 & Souza & 0.47 & Silva & 0.16 & -0.19 \\                                         
  BR & Luciano & 4458 & Silva  & 571 & 0.08 & 3.48 & 1.95 $\pm$ 0.22 & Souza  & 0.47 & Silva & 0.16 & -0.26 \\                                        
  BR & Maria & 96243 & Silva  & 11444 & 0.05 & 3.73 & 1.99 $\pm$ 0.22 & Moreira  & 0.44 & Silva & 0.16 & -0.14 \\ \hline
  PT & Ana & 3685 & Silva  & 544 & 0.14 & 3.31 & 1.98 $\pm$ 0.30 & Silva  & 0.50 & Silva & 0.13 & -0.25 \\ 
  PT & Antonio & 5708 & Pereira & 687 & 0.17 & 3.30 & 1.98 $\pm$ 0.31 & Pereira & 0.49 & Silva & 0.09 & -0.27 \\ 
  PT & Jo\~{a}o & 4639 & Santos & 687 & 0.17 & 3.26 & 2.04 $\pm$ 0.28 & Santos & 0.50 & Santos & 0.12 & -0.27 \\ 
  PT & Joaquim & 3307 & Silva & 510 & 0.16 & 3.33 & 2.00 $\pm$ 0.26 & Silva & 0.49 & Silva & 0.14 & -0.25 \\                                         
  PT & Jos\'e & 6090 & Silva & 1088 & 0.18 & 3.25 & 2.07 $\pm$ 0.24 & Silva & 0.52 & Silva & 0.16 & -0.26 \\                                         
  PT & Luiz & 3830 & Santos & 494 & 0.15 & 3.30 & 2.01 $\pm$ 0.23 & Santos & 0.49 & Santos & 0.11 & -0.28 \\                                         
  PT & Manuela & 5785 & Ferreira & 693 & 0.17 & 3.30 & 1.95 $\pm$ 0.21 & Ferreira & 0.49 & Ferreira & 0.09 & -0.28 \\ 
  PT & M\'ario & 7015 & Costa & 654 & 0.15 & 3.31 & 1.83 $\pm$ 0.38 & Costa & 0.49 & Costa & 0.06 & -0.28 \\                                         
  PT & Paulo & 3387 & Silva & 523 & 0.16 & 3.25 & 2.00 $\pm$ 0.20 & Silva & 0.51 & Silva & 0.13 & -0.28\\
\br
\end{tabular}\\
\end{table}
\normalsize
\qquad For the simulated Brazilian \textbf{NS} the value found for the parameter $\gamma$ was in the range 1.88 to 1.99. In the case of the sample containing the word Maria we have 104323 nodes and in the sample containing the word Luciano only 5814 nodes. In spite of the diferent size of the clusters, the parameter $\gamma$ for both is the order of 2, i.e., the thermodynamic limit is reached for small networks. We found the dominance of the family name \textbf{Silva} for all networks. Our model presentes a preferential attachement probablity connection but it is not a Barab\'asi like model because our model is a disconnected graph. The Brazilian society seems rather homogeneous, because few surnames belong to most of the population.\\ 
\qquad For the Portuguese \textbf{NS} the value found that the parameter $\gamma$ was in the range 1.83 to 2.07. These values agree with the Brazilian names networks. In table \ref{tab: BRPT} we observe that the surname \textit{Silva hub} appears in every Brazilian network, while in the Portuguese society depending on the group analyzed several surnames could become hub.\\    
\qquad We find that the networks studied are \textbf{small world} networks. One way to verify this feature is comparing them with random graphs \cite{ERDOS1959}.
In table \ref{tab: RSxGA} the average clustering coefficient and shortest average path are compared to random graphs (of the Erd\"{o}s-R\'enyi model - \textbf{ER model}) for the same number of nodes and adges, to be comparable.
\begin{table}
\caption{\label{tab: RSxGA}\textit{Main differences between the networks of surnames, Random Graphs (ER model), Barab\'asi-Albert model (BA) and Bianconi-Barab\'asi model (BB)}. CT means the country; \textbf{$N$}  represents the number of nodes in the network; \textbf{$C_m$} is the average clustering coefficient; \textbf{$\ell$} is the shortest average path between any two nodes in the network and $H$ is the Shannon entropy. Here we use the values of the largest subnetwork.}
\footnotesize
\centering 
\begin{tabular}{@{}lllllll}
\br
CT&network&$N$&Edges&$C_{m}$&$\ell$&$H$\\
\mr
  BR & Ana  & 22985 & 53920 & 0.07 & 3.73 & 1.20 \\
    & ER & 22985 & 53920 & $2.0\times10^{-4}$ & 6.67 & 2.17 \\
    & BA & 22985 & 53920 & $2.4\times10^{-3}$ & 5.35 & 1.72 \\
    & BB & 22985 & 53920 & $1.7\times10^{-2}$ & 4.51 & 1.50 \\ \hline
  BR & Francisco  & 13265 & 27974 & 0.058 & 3.66 & 1.06\\
    & ER & 13265 & 27974 & $2.9\times10^{-4}$ & 6.73 & 2.11\\
    & BA & 13265 & 27974 & $3.7\times10^{-3}$ & 5.13 & 1.72 \\
    & BB & 13265 & 27974 & $2.2\times10^{-2}$ & 4.37 & 1.50 \\ \hline
  BR & Jo\~{a}o  & 17341 & 43053 & 0.09 & 3.63 & 1.20 \\
    & ER & 17341 & 43053 & $2.8\times10^{-4}$ & 6.26 & 2.20\\
    & BA & 17341 & 43053 & $3.0\times10^{-3}$ & 5.24 & 1.72 \\
    & BB & 17341 & 43053 & $1.9\times10^{-2}$ & 4.44 & 1.50 \\ \hline
  BR & Jos\'e  & 38651 & 89095 & 0.08 & 3.65 & 1.21 \\ 
    & ER & 38651 & 89095 & $1.1\times10^{-4}$ & 7.07 & 2.16 \\
    & BA & 38651 & 89095 & $1.6\times10^{-3}$ & 5.55 & 1.72 \\
    & BB & 38651 & 89095 & $1.3\times10^{-2}$ & 4.64 & 1.50 \\ \hline
  BR & Maria  & 96243 & 248424 & 0.05 & 3.73 & 1.46 \\
    & ER & 96243 & 248424 & $5.0\times10^{-5}$ & 7.17 & 2.22 \\ 
    & BA & 96243 & 248424 & $7.4\times10^{-4}$ & 5.89 & 1.72 \\
    & BB & 96243 & 248424 & $8.4\times10^{-3}$ & 4.86 & 1.50 \\ \hline
  PT & Antonio  & 5708 & 22257 & 0.17 & 3.30 & 1.97 \\
    & ER & 5708 & 22257 & $1.3\times10^{-3}$ & 4.44 & 2.43 \\
    & BA & 5708 & 22257 & $8.6\times10^{-3}$ & 4.08 & 2.06 \\
    & BB & 5708 & 22257 & $3.8\times10^{-2}$ & 3.58 & 1.83 \\ \hline
  PT & Jo\~{a}o & 4639 & 17618 & 0.18 & 3.26 & 1.96 \\ 
    & ER & 4639 & 17618 & $1.6\times10^{-3}$ & 4.39 & 2.42\\
    & BA & 4639 & 17618 & $1.0\times10^{-2}$ & 4.01 & 2.06 \\
    & BB & 4639 & 17618 & $4.2\times10^{-2}$ & 3.54 & 1.83 \\ \hline
  PT & Jos\'e & 6090 & 24182 & 0.18 & 3.25 & 1.98 \\ 
    & ER & 6090 & 24182 & $1.3\times10^{-3}$ & 4.44 & 2.44 \\                                         
    & BA & 6090 & 24182 & $8.2\times10^{-3}$ & 4.11 & 2.06 \\
    & BB & 6090 & 24182 & $3.7\times10^{-2}$ & 3.60 & 1.82 \\ \hline
  PT & Manuela & 5785 & 23553 & 0.17 & 3.30 & 1.98 \\
    & ER & 5785 & 23553 & $1.4\times10^{-3}$ & 4.37 & 2.45 \\
    & BA & 5785 & 23553 & $9.9\times10^{-3}$ & 3.72 & 2.31 \\ 
    & BB & 5785 & 23553 & $4.2\times10^{-2}$ & 3.30 & 2.06 \\  \hline
  PT & M\'ario & 7015 & 27428 & 0.15 & 3.31 & 1.84 \\
    & ER & 7015 & 27428 & $1.1\times10^{-3}$ & 4.54 & 2.43 \\
    & BA & 7015 & 27428 & $7.3\times10^{-3}$ & 4.15 & 2.06 \\
    & BB & 7015 & 27428 & $3.4\times10^{-2}$ & 3.64 & 1.82 \\  
\br
\end{tabular}\\
\end{table}
\normalsize
\subsection{Centrality}
\begin{figure*}[htb!]
\centering
\begin{minipage}{0.5\linewidth}
\subfigure[Luciano]{\includegraphics[scale=0.39]{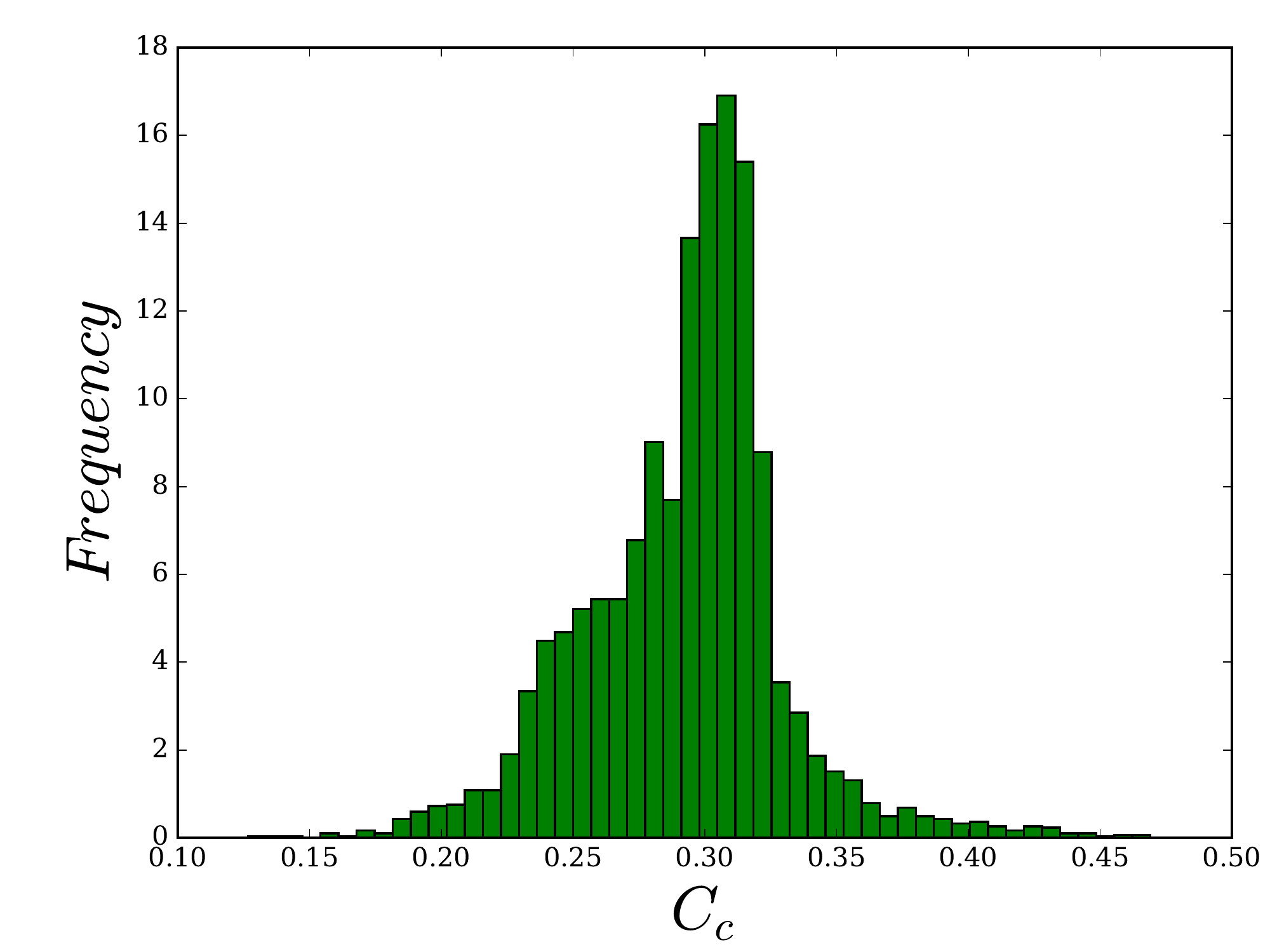}}
\end{minipage}\hfill
\begin{minipage}{0.5\linewidth}
\subfigure[Ana]{\includegraphics[scale=0.40]{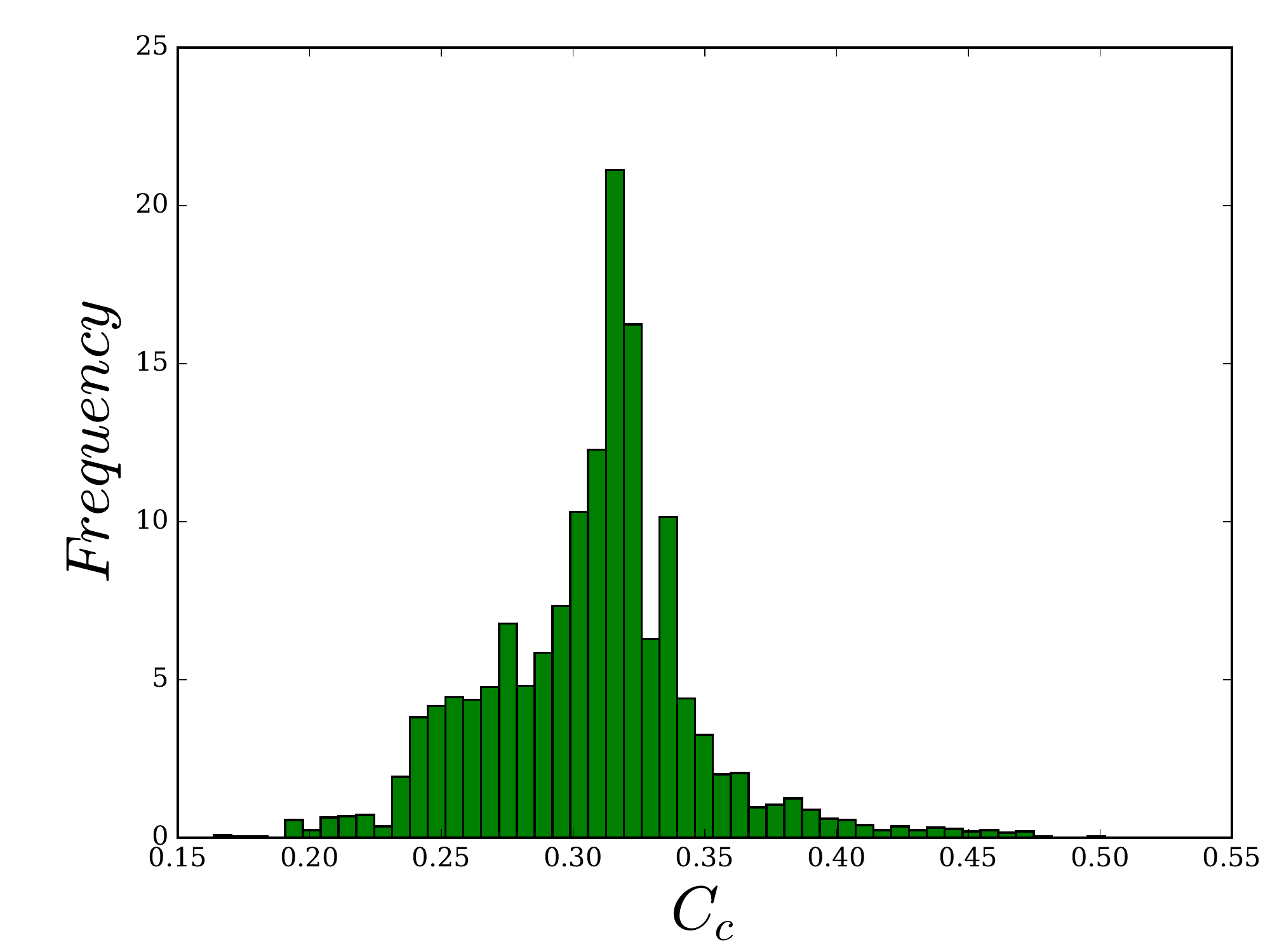}}
\end{minipage}
\caption{\small (a) Histogram of the closeness centrality for a sample with 4458 nodes of the Brazilian population whose names contain the word \textbf{Luciano} and (b) a sample with 3685 nodes of the Portuguese population whose names contain the word \textbf{Ana}. The plot shows the number of nodes vs. the closeness centrality.}
\label{C_central}
\end{figure*}

\begin{figure*}[htb!]
\begin{center}
\includegraphics[scale=0.45]{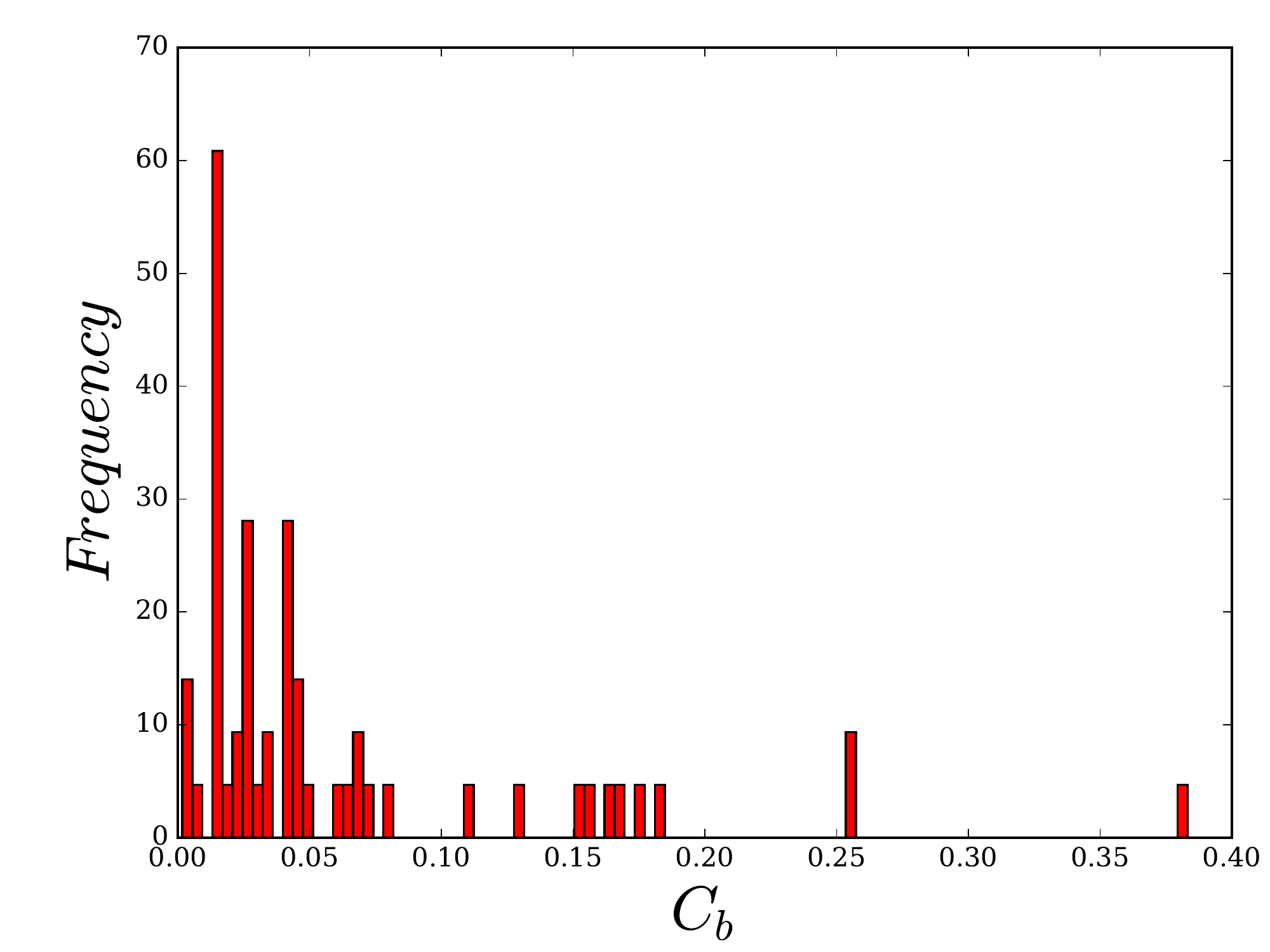}
\caption{\small Histogram of the betweenness centrality for a list of names 250 containing the word \textbf{Jos\'e}. The plot shows the frequency appearance vs. the betweenness centrality. The betweenness centrality of vertices is very low, most have zero centrality and in general the most central node is the main hub of the network.}
\label{B_central}
\end{center}
\end{figure*}

\qquad The \textbf{Closeness centrality} $C_c$ is defined by:
\begin{equation}
C_{c_i} = \sum_{j}[d_{ij}]^{-1} =  \frac{1}{\sum_{j}d_{ij}}.
\label{eq-closeness-centrality} 
\end{equation}
\qquad High centrality indices indicate that a node can reach other nodes on a relatively short path, or a node is connected to others by considerable fractions of shortest paths \cite{BRANDES2001}. We have studied the closeness centrality of \textbf{NS} to identify which family name is closest to all others family names since we believe that an important node must be a very short distance from any other randomly chosen node in the network.\\
\qquad One obstacle to calculate the closeness centrality of \textbf{NS} is the fact that it is not a connected graph, because our social networks are often disconnected \cite{MOXLEY1974}. The alternative is to rewrite the closeness equation as the sum of inverse distances to all other nodes rather than the inverse of the sum of distances to all other nodes \cite{OPSAHL2010} and \cite{NEWMAN2003b}: 

\begin{equation}
C_{c_i} = \sum_{j}\frac{1}{d_{ij}}.
\label{eq-closeness-centrality-new}
\end{equation} 
This is also called \textit{harmonic centrality} \cite{NEWMAN2003b}.\\
\qquad The vertex with higher closeness centrality of the Brazilian networks surprisingly does not coincide with the hub of the network. Differently, the vertex with higher closeness centrality of the Portuguese networks coincides with the main hub of the network. In Fig. \ref{C_central} we show the histogram of the closeness centrality for (a) a sample of Brazilian society and (b) a sample of Portuguese society.\\
\qquad Another important centrality is the \textbf{Betweenness centrality} $C_b$. In \cite{FREEMAN1977} the betweenness centrality is presented as an alternative to the centrality calculation for disconnected graphs. Here we are interested in the most influential vertex, since this centrality gives us the number of times that a vertex belongs to all shortest path between any two vertices (optimal paths). For example, an airport that belongs to several routes, obviously is a strategic point. Thus we want to know which family name belongs to the greatest number of optimal paths between two randomly selected vertices in the network.\\
\qquad For a given graph $G=(V,E)$ with $V$ nodes, the betweenness centrality is defined by:
\begin{equation}
C_{b_i} = \sum_{j\neq i\neq k\epsilon V}\frac{\sigma_{jk}(i)}{\sigma_{jk}},
\label{eq-betweenness-centrality}
\end{equation}
where $\sigma_{jk}$ is total number of shortest paths from node $j$ to node $k$ and $\sigma_{jk}(i)$ is the number of those paths that pass through $i$.
For both studied societies, the vertex with higher betweenness centrality of the \textbf{NS} generally coincides with hub of the network. In Fig. \ref{B_central} we show the histogram of the betweenness centrality for a sample of Portuguese society.


\subsection{Assortative mixing}
\qquad The assortative mixing measures the tendency of nodes to associate with nodes of similar degree as follows:
\begin{equation}
r = \frac{M^{-1}\sum_{i}j_ik_i-[M^{-1}\sum_{i}\frac{1}{2}(j_i+k_i)^2]}{M^{-1}\sum_{i}\frac{1}{2}(j_i^2+k_i^2)-[M^{-1}\sum_{i}\frac{1}{2}(j_i+k_i)^2]},
\label{eq-assortative-mixing}
\end{equation}
where $M$ is the total number of edges, $j_i$ and $k_i$ are the degree of the nodes that are on both ends of the i-th edge \cite{NEWMAN2002}.\\
\qquad All surname networks in this study have disassortative mixing, i.e., high degree vertices preferentially connect with low degree ones and vice versa.\\

\subsection{Shannon Entropy}
\qquad We calculate the Shannon entropy \cite{SHANNON1948} for all networks in order to quantify the disorder of the system. The Shannon entropy is defined as:
\begin{equation}
H = -K \sum_i \left( p_i \log p_i \right),
\label{eq-shannon-entropy}
\end{equation}
in this work we used $K=1$.\\ 
\qquad In table \ref{tab: BRPT} values are displayed for the closeness centrality, betweenness centrality, assortative mixing and in table \ref{tab: RSxGA} we show the Shannon entropy for the \textbf{NS} Brazilian and Portuguese, random graphs (Erd\"{o}s-R\'enyi model), Barab\'asi-Albert model and Bianconi-Barab\'asi model. From the centrality and entropy results we can conclude that the Portuguese society is more heterogeneous than the Brazilian society.


\section{Conclusion and discussion}
\qquad In this work we analyze the surname networks of two distinct societies (Brazilian and Portuguese). We start from telephone listings and make a survey of the family names existing in these societies. From these surnames we model the network by establishing the edges (individuals) that connect two family names and obtain a disconnected graph. With these networks we study the following properties: connectivity distribution, shortest average path, clustering coefficient, closeness centrality and betweenness centrality, degree correlation and entropy of degree distribution. As a result, we verified that all networks studied here have scale-free properties with a characteristic exponent $\gamma \sim$ 2. All Brazilian networks studied have the family \textbf{Silva} as the main hub, while Portuguese networks have a variety of family names as main hub of their networks. We study the centralities to determine the most important families in these networks, we use the closeness centrality to determine the vertices that are closest to any other vertex of the network and the betweenness centrality to determine the most influential vertices. The vertex with higher closeness centrality of the Brazilian networks generally does not coincide with the hub of the network, while in the Portuguese networks these vertices generally coincide. The vertices with the higher betweenness centrality, that is, those belonging to optimal paths, usually coincide with the hubs of the networks in both societies studied. We have shown that, although Brazilian and Portuguese societies are known to share similarities, yet the family name network is more homogeneous for Brasil than for Portugal. In contrast, the Portuguese society is more clustered as far as surnames are concerned, and this society has a higher entropy for the degree distribution what characterizes a less structured network (more random). Remarkably, the surname networks are dissassortative, unlike most social networks. Finally, as expected, we find that the networks possess the small world characteristic. Its interesting to study others countries with different languages (different cultures) to look for really universal properties.

\section*{Acknowledgments}
\qquad We thank the financial support of CAPES/CNPq, Thiago Cris\'ostomo, Samura\'i Brito, Ricardo Borges and Tiago Medeiros for the discussions that contributed to the quality of this work.
 


\section*{References}

\begin{thebibliography}{}
\bibitem{GARTON1997} Garton L, Haythornthwaite C and Wellman B Studying online social networks \textit{Journal of Computer-mediated Communication} \textbf{3} 1997

\bibitem{HANNEMAN2001} Hanneman R A Introduction to social networks methods \textit{University of California, UC} \textbf{2001}
  
\bibitem{DURKHEIN1995} Durkhein \'E As Regras do M\'etodo Sociol\'ogico \textit{Livraria Martins Fontes Editora Ltda, SP-BR} \textbf{1995}  
  
\bibitem{MORENO1934} Moreno J L Who Shall Survive? \textit{Beacon House, NY} \textbf{1934}  

\bibitem{HARARY1969} Harary F Graph Theory \textit{Addison-Wesley} \textbf{1969}  

\bibitem{DOROGOVTSEV2002} Dorogovtsev S N  and Mendes J F F Evolution of networks \textit{Adv. in Phys.} \textbf{51} 1079-1187 2002

\bibitem{LIND2007} Lind P G, da Silva L R, Andrade Jr. J S and Herrmann H J The Spread of Gossip in American Schools \textit{Europhysics Letters} \textbf{78} \textbf{p.68005} 2007
  
\bibitem{LIND2007A} Lind P G, da Silva L R, Andrade Jr. J S and Herrmann H J Spreading Gossip in Social Networks \textit{Physical Review E} \textbf{76} \textbf{p.036117} 2007

\bibitem{MIYAZIMA2000} Miyazima S, Lee Y, Nagamine T and Miyajima H Power-Law Distribution of family names in Japanese Societies \textit{Physica A} \textbf{278} 282-288 2000

\bibitem{NEWMAN2010} Newman M E J Networks: An Introduction \textit{Oxford  University Press, Oxford} \textbf{2010} 

\bibitem{BARABASI2002} Barab\'asi A-L, Jeong H, N\'eda Z, Ravasz E, Schubert A and Vicsek T Evolution of the Social Network of Scientific Collaborations \textit{Physica A} \textbf{311} 590-614 2002

\bibitem{NEWMAN2003a} Newman M E J Mixing Patterns in Networks \textit{Physical Review E} \textbf{67} 026126-1-026126-13 2003

\bibitem{NEWMAN2003b} Newman M E J The Structure and Function of the Complex Network \textit{SIAM Review} \textbf{45} 167-256 2003

\bibitem{CLAUSET2009} Clauset A, Shalizi C R and Newman M E J Power-law distribuitions in empirical data \textit{SIAM Review} \textbf{51} 2009

\bibitem{PRICE1965} Price D J de S Networks of scientific papers \textit{Science} \textbf{149} 510-515 1965
 
\bibitem{BARABASI1999} Barab\'asi  A-L and Albert R Emergence of scaling in random networks \textit{Science} \textbf{286} 509-512 1999

\bibitem{BARABASI_ALBERT2002} Barab\'asi A-L and Albert R Statistical mechanics of complex networks \textit{Reviews of Modern Physics} \textbf{74} 2002

\bibitem{BARABASI2001a} Barab\'asi  A-L and Bianconi G Competition and multiscaling in evolving networks \textit{Europhys. Lett} \textbf{54} 436-442 2001

\bibitem{ERDOS1959} Erd\"{o}s P and R\'enyi A On random graphs \textit{Publ. Math} \textbf{6} 290-297 1959

\bibitem{MOXLEY1974} Moxley R L and Moxley N F Determining point-centrality in uncontrived social networks \textit{Sociometry} \textbf{37} 122-130 1974

\bibitem{BRANDES2001} Brandes U A faster algorithm for betweenness centrality \textit{The Journal of Mathematical Sociology} \textbf{25(2)} 2001
  
\bibitem{OPSAHL2010} Opsahl T, Agneessens F and Skvoretz J Node centrality in weighted networks: Generalizing degree and shortest paths \textit{Social Networks} \textbf{32(3)} 245-251 2010

\bibitem{FREEMAN1977} Freeman L C A set of measures of centrality based on betweenness \textit{Sociometry} \textbf{40} 35-41 1977
   
\bibitem{NEWMAN2002} Newman M E J Assortative Mixing in Networks \textit{PRL} \textbf{89} \textbf{p.208701} 2002

\bibitem{SHANNON1948} Shannon C A mathematical theory of communication \textit{Bell System Technology Journal} \textbf{27} 379-423 1948
  

\end{thebibliography}
\end{document}